\documentclass [twocolumn,aps,prl]{revtex4}
\usepackage{epsfig,graphicx,amssymb,color,amssymb,amsmath}

\begin{document}
\title{Numerical Study of Carrier Multiplication Pathways in Nanocrystalline and Bulk Form of PbSe}
\author{Kirill A. Velizhanin}
\author{Andrei Piryatinski}\email{apiryat@lanl.gov}

\address{Center for Nonlinear Studies (CNLS),
 Theoretical Division, Los Alamos National Laboratory, Los Alamos, NM 87545 }

\date{\today}

\begin{abstract}
Employing the interband exciton scattering model, we have performed a numerical study of the direct photogeneration and population relaxation processes contributing to carrier multiplication (CM) in nanocrystalline and bulk PbSe. We argue that in {\it both} cases the impact ionization is the main mechanisms of CM. This explains weak contribution of the direct photogeneration to the total quantum efficiency (QE). Investigated size scaling of QE in NCs and comparison to the bulk limit provide microscopic insight in the experimentally observed trends. 
\end{abstract}
\maketitle


Recent demand in new efficient photovoltaic devices has turned significant attention to the problem of carrier multiplication (CM) in semiconductor materials~\cite{kolodinski93,nozik02}. Early reported ultrafast spectroscopic studies of nanocrystals (NCs)~\cite{schaller04,ellingson05,schaller05} suggested that nanostructuring is a direct way to highly increase the CM quantum efficiency (QE), i.e., number of excitons produced per absorbed photon. This was supported by the following arguments: An enhancement of the Coulomb interaction between carriers due to their spatial confinement should lead to more efficient biexciton production. Relaxation of the quasimomentum conservation constraint by breaking the translational symmetry should open additional pathways for CM while the intraband cooling is slowed  down by the presence of the phonon bottleneck. More recent reports claim much lower QE or the absence of CM in NCs~\cite{nair07,nair08,benlulu08,pijpers08}. The controversy possibly rise from experimental inaccuracy~\cite{benlulu08}, sample-to-sample variation in surface properties~\cite{beard09}, and contribution from extraneous effects, e.g., photocharging~\cite{mcguire08,mcguire10}. Comparison to the bulk PbS and PbSe spectroscopic measurements shows that bulk QE exceeds validated QE in NCs if compared on the scale of absolute energy~\cite{pijpers09}. These issues call for the reassessment of the quantum confinement role in the CM dynamics~\cite{nair08}.  

From theory point of view, multiexciton photogeneration~\cite{schaller05,rupasov07}, coherent multiexciton production~\cite{shabaev06,witzel10}, and incoherent impact ionization (II)~\cite{franceschetti06,rabani08,allan08},  are currently under debate as primary mechanisms of CM. The {\it ab initio} calculations on small {\it clusters} ($\lesssim 1$~nm)  demonstrated the role of strong Coulomb correlations in multiexciton photogeneration and emphasized the role of fast exciton-biexciton dephasing~\cite{prezhdo09}. Atomistic calculations focused on small diameter ($\lesssim 3$~nm) NCs show that CM is dominated by II processes~\cite{allan08,rabani10}. An extrapolation of these calculations to larger NCs and comparison to the bulk have been reported recently~\cite{delerue10}. Effective mass models have been used to study direct photogeneration~\cite{silvestri10} and coherent CM dynamics in larger ($4-6$~nm) diameter NCs. Reported experimental studies consider even larger diameter ($\lesssim 10$~nm) NCs. However, no systematic investigation of {\em all} predicted CM pathways contributing to the ultrafast spectroscopic signal (i.e., photogeneration and population relaxation) in  various size NCs and comparison to the bulk limit has been reported yet.      

In this letter, we report on such study employing the interband exciton scattering model (IESM)~\cite{piryatinski10} parameterized for PbSe materials. The IESM treats on equal footing the biexciton photogeneration during interaction with the pump pulse and subsequent population relaxation processes. This allows us to evaluate the contributions of all predicted pathways to the total QE. To clarify the quantum confinement effect on the key quantities determining QE, comparison with the bulk limit is reported on the absolute energy scale~\cite{nair08,delerue10}. In general, performance of the photovoltaic devices can be characterized by variously defined power efficiency~\cite{mcguire10,delerue10}, in particular using the photon energy scale {\it normalized} per NC/bulk band gap energy~\cite{beard10}. Since we focus on fundamental mechanisms of CM rather than applications, the latter unitless scale is not used below. 

We consider an ensemble of NCs in which no more than one photon can be absorbed per NC leading to production of either exciton or biexciton state. Hence, QE can be defined as 
\begin{eqnarray}
\label{QE}
    QE =(2N_{xx}+N_x)/(N_{xx}+N_x),
\end{eqnarray}
where $N_x$ and $N_{xx}$ are the ensemble non-equilibrium exciton and biexciton populations, respectively. 
For an ensemble of NCs, the bulk limit can be defined as the thermodynamic limit: $V\rightarrow\infty$, $V/v\rightarrow\infty$, and $v=const$. Here, $V$ is a NC volume, $v$ is the unit cell volume, and ratio $V/v$ gives the number of unit cells. Since, QE (Eq.~(\ref{QE})) is volume independent in the bulk limit, it is referred to as the {\em intensive} parameter. As a result, its scaling with NC size compared to the bulk limiting value provides convenient measure of the quantum size effects. In general, the populations $N_x$ and $N_{xx}$ as well as DOS and Coulomb interactions (determining $N_x$ and $N_{xx}$) are not intensive quantities. To make proper comparison with the bulk limit, below, we eliminate their volume scaling by multiplying them with powers of $(v/V)$. 

We start by defining the populations $n^{x}_a$ and $n^{xx}_k$ of $a$-th exciton and $k$-th biexciton states, respectively. Their time evolution during the interaction with the pump pulse (photogeneration) and subsequent population relaxation are numerically calculated using the weak Coulomb limit of IESM~[\onlinecite{piryatinski10}]. To account for high DOS region in which CM dynamics occur, we recast $n^{x}_a$, and $n^{xx}_k$ to the quasicontinuous frequency representation: $n_x(\omega)=\sum_{a\geq1} n^{x}_a\delta(\omega-\omega^{x}_a)$  and $n_{xx}(\omega)=\sum_{k\geq1} n^{xx}_k\delta(\omega-\omega^{xx}_k)$. Accordingly, the {\em intensive} populations entering Eq.~(\ref{QE}) become defined as $N_x=(v/V)\int_0^\infty d\omega~n_x(\omega)$ and $N_{xx}=(v/V)\int_0^\infty d\omega~n_{xx}(\omega)$. 

In most ultrafast CM experiments, the pump duration exceeds typical dephasing time leading to the continuous wave (CW) excitation regime. Our simulations show that in agreement with Ref.~\cite{witzel10}, high DOS leads to the suppression of the oscillating terms that contribute to $n^{x}_a$ and $n^{xx}_k$. The IESM predicts interference of the photogeneration pathways contributing to $n^{xx}_k$~\cite{piryatinski10}. However, the simulations show that due to high DOS, the sign-varying terms cancel out fully eliminating the interference. Taking all this into account, the CW photogenerated biexciton population can be represented as
\begin{eqnarray}\label{NxxE}
&~&N_{xx}(\omega_{pm})=
\\\nonumber&~&
\frac{\cal A}{\hbar^2}\int d\omega^{'}[V^{x,xx}_{eff}(\omega_{pm},\omega^{'})]^2
	\frac{\tilde\rho_{x}(\omega_{pm})\rho_{xx}(\omega^{'})}{(\omega^{'}-\omega_{pm})^2+\gamma^2}
\\\nonumber&+&	
	\frac{\cal A}{\hbar^2}\int d\omega^{'}[V^{x,xx}_{eff}(\omega^{'},\omega_{pm})]^2
	\frac{\tilde\rho_{x}(\omega^{'})\rho_{xx}(\omega_{pm})}{(\omega^{'}-\omega_{pm})^2+\gamma^2}
\\\nonumber&+&	
\frac{A}{\hbar^2}\int d\omega^{'}[V^{xx}_{eff}(\omega^{'})]^2
	\frac{\tilde\rho_{xx}(\omega^{'}\omega_{pm})}{{\omega^{'}}^2},
\end{eqnarray}
and the leading contribution to the exciton population as $N_{x}(\omega_{pm})={\cal A}\tilde\rho_x(\omega_{pm})$. Here, ${\cal A}$ is proportional to the pump fluence, $\omega_{pm}$ is the pump frequency, and $\gamma$ is the dephasing rate between exciton and biexciton states. $\rho_{x}(\omega)=(v/V)^2\sum_a\delta(\omega-\omega^x_a)$ and $\rho_{xx}(\omega)=(v/V)^4\sum_k\delta(\omega-\omega^{xx}_k)$ are {\em intensive} exciton and biexciton DOS, respectively. The associated optically allowed exciton DOS is $\tilde\rho_{x}(\omega)=(v/V)\sum_a|\mu^x_{a}|^2\delta(\omega-\omega^x_a)$ and joint biexciton DOS is $\tilde\rho_{xx}(\omega_1,\omega_2)=(v/V)^4\sum_{kl}|\mu^{xx}_{kl}|^2\delta(\omega_1-\omega^{xx}_k)\delta(\omega_2-\omega^{xx}_l)$. They depend on the interband exciton, $\mu^x_{a}$, and the intraband biexciton, $\mu^{xx}_{kl}$, transition dipoles. 
  
Key {\em intensive} quantity in Eq.~(\ref{NxxE}) is the effective Coulomb interaction which we define as the r.m.s. of the interband Coulomb matrix elements connecting the states with frequencies falling into $[\omega_1,\omega_1+d\omega_1]$ and $[\omega_2,\omega_2+d\omega_2]$ intervals,
\begin{eqnarray}\label{Veff}
	V^{x,xx}_{eff}(\omega_1,\omega_2)=\left(\frac{V}{v}\right)^2\left[\sum_{a,m}\left|V^{x,xx}_{a,m}\right|^2
\right.&~&\\\nonumber\left.\times
	\frac{\delta(\omega_1-\omega^x_a)\delta(\omega_2-\omega^{xx}_m)}
			{\sum_{b}\delta(\omega_1-\omega^x_b)\sum_{n}\delta(\omega_2-\omega^{xx}_n)}\right]^{1/2} &.&
\end{eqnarray}
Here, $\omega^x_a$  ($\omega^{xx}_m$) is $a$-th exciton ($m$-th biexciton) state frequency, and $V^{x,xx}_{a,m}$ is the Coulomb matrix element between these states. The size  scaling of $V^{x,xx}_{eff}$ in NCs reflects the net result of the Coulomb matrix elements scaling, relaxation of the momentum conservation constraints and the appearance of new selection rules associated with the confinment potential symmetry. In Eq.~(\ref{NxxE}), $V^{xx}_{eff}(\omega)$ denotes effective Coulomb coupling between the ground and biexciton states.

Three terms in the r.h.s. of Eq.~(\ref{NxxE}) describe the associated photogeneration pathways~\cite{piryatinski10}: In the first term  the optically allowed exciton DOS, $\tilde\rho_x(\omega_{pm})$, depends on the pump frequency indicating resonant production of single excitons subsequently scattered to the biexciton states. This is the {\it indirect} biexciton photogeneration pathway predicted by IESM~\cite{piryatinski10}. The second term contains biexciton DOS, $\rho_{xx}(\omega_{pm})$, depending on the pump frequency indicating the {\it direct} biexciton photogeneration through virtual exciton states. Finally, the last term describes the pathway involving direct biexciton generation by optical stabilization of the scattering processes between the ground state and biexciton manifold. The last two pathways have been first considered in Refs.~[\onlinecite{schaller05}] and [\onlinecite{rupasov07}], respectively.

The population relaxation dynamics is described by a set of kinetic equations introduced in Ref.~[\onlinecite{piryatinski10}] and further transformed to the quasicontinuous frequency representation. They contain II and Auger recombination rates
\begin{eqnarray}\label{kII}
k_{II}(\omega)&=&\frac{2\pi}{\hbar^2}[V^{x,xx}_{eff}(\omega)]^2\rho_{xx}(\omega),
\\\label{kAR}
k_{AR}(\omega)&=&\frac{2\pi}{\hbar^2}\left(\frac{v}{V}\right)^2[V^{x,xx}_{eff}(\omega)]^2\rho_{x}(\omega),
\end{eqnarray}
respectively, with $V^{x,xx}_{eff}(\omega)\equiv V^{x,xx}_{eff}(\omega,\omega)$. According to Eq.~(\ref{kII}), the II rate depends on intensive parameters and therefore has finite value in the bulk limit. In contrast, the Auger rate has uncompensated volume prefactor and vanishes as $V^{-2}$ in the bulk limit. The intraband cooling rates in the kinetic equations are calculated using Ohmic spectral density assuming no phonon bottleneck present in NCs. The total QE is evaluated through numerical solution of the kinetic equations with photogenerated populations used as the initial conditions. 

\begin{figure}[t]
\begin{center}
\epsfig{file=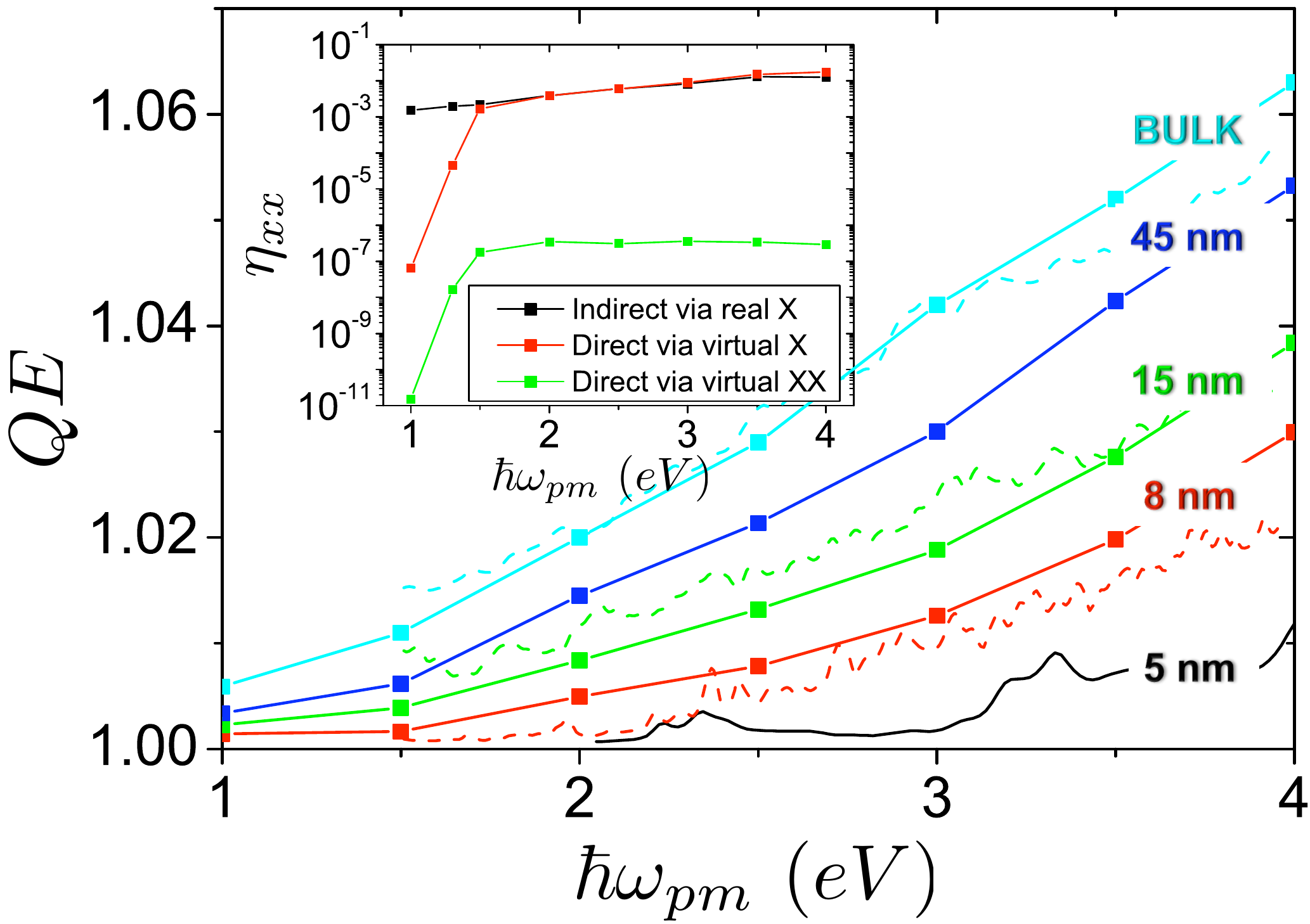,width=2.5in}
\end{center}
\caption{Photogeneration QE vs pump energy for PbSe NCs of diameter $d=5,8,15,45$~nm and bulk.
		Solid lines mark QE calculated using general expressions for populations from Ref.~\cite{piryatinski10}. Dash marks 
		approximate results (d=8, 15~nm NCs and bulk) due to Eq.~(\ref{NxxE1}). The inset shows biexciton 
		  quantum yield associated with each photogeneration pathway for $d=15$~nm.}
\label{fig-pg}
\end{figure}

To calculate the electron and hole wavefunctions in spherically symmetric PbSe NCs and bulk, we use the effective mass formalism due to Kang and Wise~\cite{kang97}. The exciton and biexciton wavefunctions are introduced as uncorrelated configurations of the electron and hole wavefunctions to evaluate the transition dipoles and the interband Coulomb matrix elements (see Supplemental Material). The exciton (biexciton) dephasing rate is set to $50$~meV (100~meV) which is order of magnitude consistent with Ref.~\cite{prezhdo09}. To account for degeneracy of the exciton (biexciton) states associated with four equivalent $L$-valleys, the exciton (biexciton) population is multiplied by factor 4 (16). Introducing energy independent effective electron-phonon coupling in the high DOS region, the intraband cooling rates are obtained by fitting the total relaxation time, $\tau_{ph}$, to reproduce the experimentally observed values in the range  $0.5<\tau_{ph}<5.0$~ps~\cite{allan08,pijpers09}. The Monte Carlo sampling is used to handle ensemble of NCs with  $5\%$ size distribution. 

\begin{figure}[t]
\begin{center}
\epsfig{file=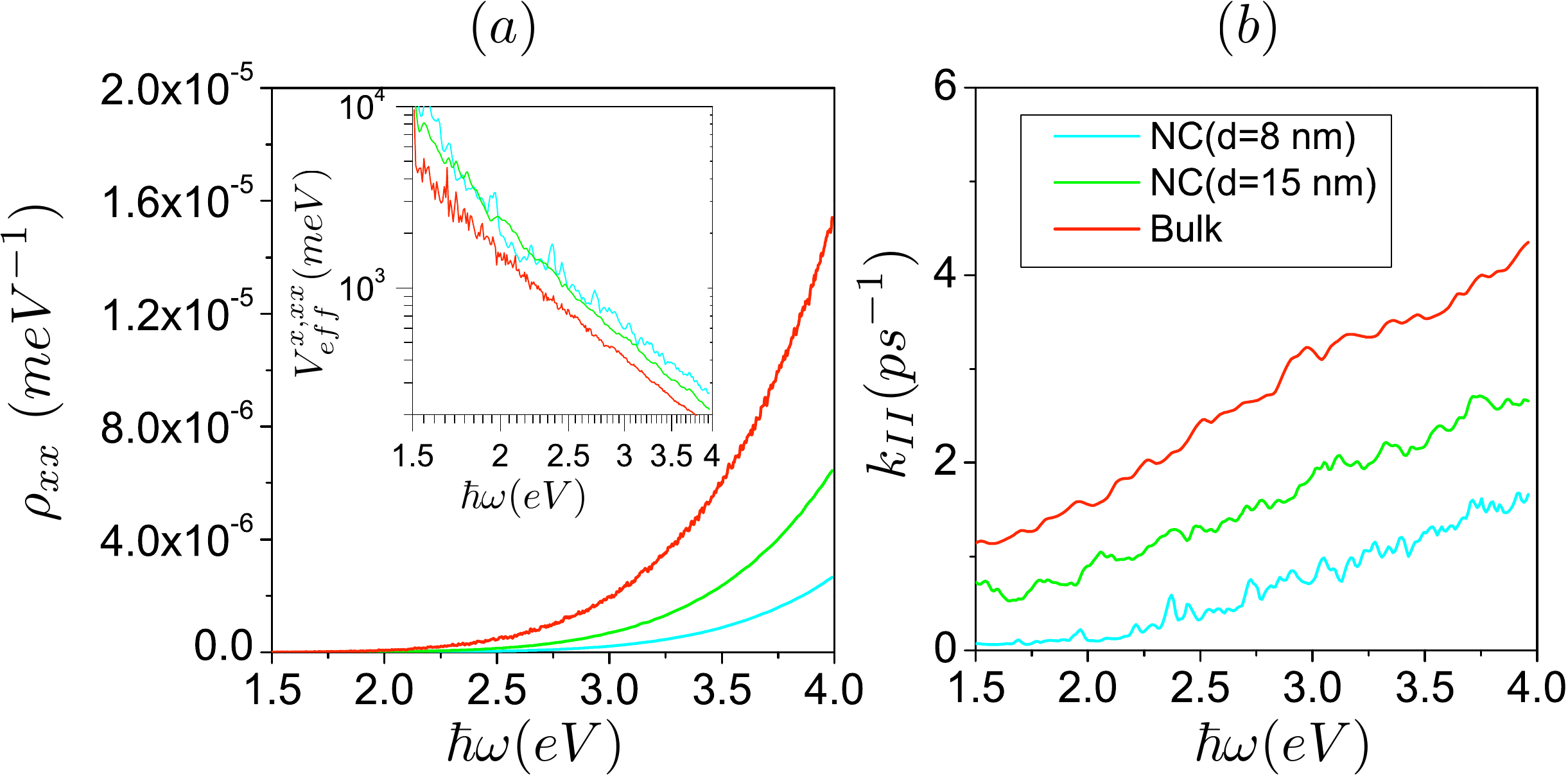,width=3.5in}
\end{center}
\caption{Calculated (a) intensive biexciton DOS, (inset) effective Coulomb interaction, and (b) II rate vs energy for PbSe NCs of diameter $d=8,15$~nm and bulk.}
\label{fig-ii}
\end{figure}

Fig.~\ref{fig-pg} shows monotonic increase of the calculated photogeneration QE in the NC ensemble to the bulk values as the NC mean diameter increases. To explain this trent, we first look at the contributions of three photogeneration pathways (Eq.~(\ref{NxxE})) to the biexciton quantum yield, $\eta_{xx}=QE-1$. According to the inset to Fig.~\ref{fig-pg}, the contribution of the last pathway (green line) is negligible. The other two become equal above the photogeneration energy threshold, $\sim 1.5$~eV. Taking into account resonant nature of the denominator in the first two terms of Eq.~(\ref{NxxE}), the integral convolutions can be calculated. This results in the first and second term identical contributions to $\eta_{xx}$ as observed in Fig.~\ref{fig-pg}. Hence, the total  biexciton population acquires very simple dependence
\begin{eqnarray}\label{NxxE1}
N_{xx}(\omega_{pm}) = k_{II}(\omega_{pm})N_{x}(\omega_{pm})/\gamma_{eff},
\end{eqnarray}
on the II rate, $k_{II}$, photogenerated exciton population, $N_{x}$, and the effective exciton-biexciton dephasing rate, $\gamma_{eff}$. As a result, the photogeneration QE becomes dependent only on the II rate and effective dephasing rate, $QE = 1+k_{II}/\gamma_{eff}$. To check this relation, we calculated QE using II rate described by Eq.~(\ref{kII}) (Fig.~\ref{fig-ii}) with $\hbar\gamma_{eff}=50$~meV. According to Fig.~\ref{fig-pg}, the latter QE (dashed lines), well reproduces the trends of the ``exact" calculations (solid lines). 

The analysis above leads to an important conclusion that in the absence of interference destroyed by high DOS and the interband dephasing processes, {\it the photogeneration becomes resonant process}. Furthermore, the biexciton generation described by Eq.~(\ref{NxxE1}) can be interpreted as a {\it single II event} of the exciton population occurring on the dephasing timescale, $\gamma_{eff}^{-1}$. Previous studies of photogeneration processes~\cite{schaller05,rupasov07}, implicitly assumed {\em constructive} interference {\it within} the second and the third pathways. The assumption led to significant overestimation of photogeneration QE. Established linear dependence between the photogenerated QE and II rate (Fig.~\ref{fig-ii}(b)) is a key to explain the QE  variation with NC size observed in Fig.~\ref{fig-pg}. According to the inset to Fig.~\ref{fig-ii}(a), the effective Coulomb interaction (Eq.~(\ref{Veff})) entering II rate (Eq.~(\ref{kII})) is  enhanced in NCs compared to the bulk limit just  by a factor of two. In contrast, our calculations show that the biexciton DOS (Fig.~\ref{fig-ii}(a)) reduction in NCs due to quantum confinement fully overplays the effective Coulomb enhancement. The dominant role of the DOS in II rate leads to the monotonic increase of QE in NC with their size increasing as observed in Fig.~\ref{fig-pg}. This has also been found in Ref.~\cite{pijpers09}

\begin{figure}[t]
\begin{center}
\epsfig{file=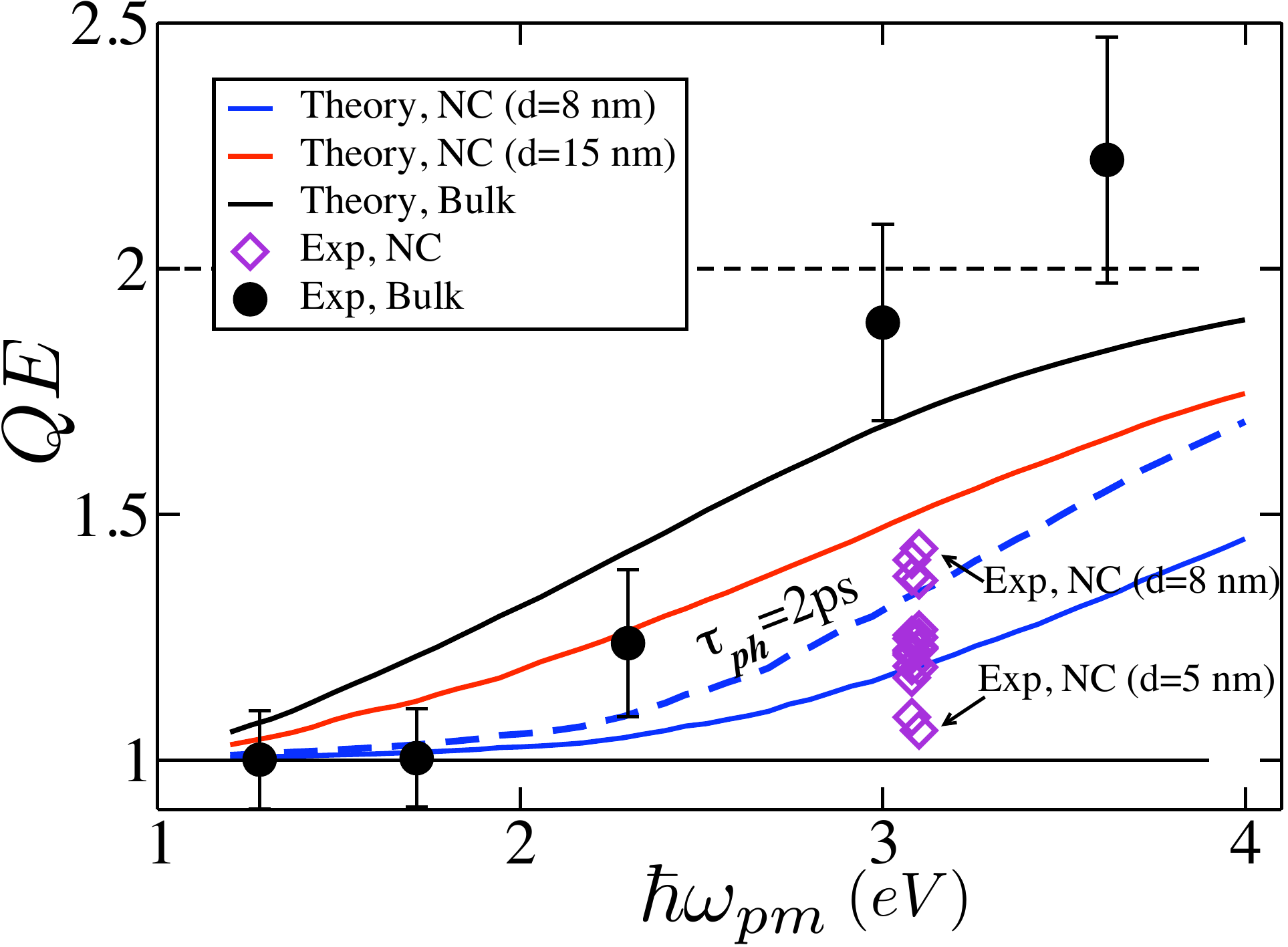,width=2.4in}
\end{center}
\caption{Total QE as a function of pump energy calculated for PbSe NCs of diameter $d=8,15$~nm and bulk. 
		 Solid (dashed) lines present calculations with $\tau_{ph}=1$~ps ($\tau_{ph}=2$~ps). 
		 For comparison, diamonds and dots mark experimental data from Refs.~[\onlinecite{mcguire08}] 
		 and [\onlinecite{pijpers09}], respectively.}
\label{fig-pr}
\end{figure}

In contrast to II rate (Eq.~(\ref{kII})), the volume scaling of Auger recombination rate (Eq.~(\ref{kAR})) makes this competing process negligible as the NC diameter increases. Therefore, the QE associated with the population relaxation is fully determined by the interplay between comparable II and cooling, $\tau_{ph}$ times. As a result, the {\it total} (i.e., photogeneration and population relaxation) QE has the same trend with the NCs size variation (Fig.~\ref{fig-pr}) as  II rate  (Fig~\ref{fig-ii}(b)) and never exceeds the bulk values. For comparison, the experimental data for PbSe NCs~\cite{mcguire08} and the bulk~\cite{pijpers09} are also shown in the plot. Solid lines in Fig.~\ref{fig-pr} show QE calculated for the intraband cooling time $\tau_{ph}=1.0$~ps which provides the best fit for the bulk.\cite{pijpers09} Generally, the theory reproduces the experimental trends in bulk (also supporting the adopted model for the phonon assisted cooling). Note that for $\hbar\omega_{ph}\gtrsim 3.6$~eV where the tri-exciton generation appears, the application of our model becomes limited. To get better agreement with experiment for NC of $d=8$~nm, we had to increase the cooling time up to $\tau_{ph}=2$~ps (blue dash). The increase of $\tau_{ph}$ can be attributed to the quantum confinement induced increase in the level spacing~\cite{bonati07}.  The photogeneration contribution (Fig.~\ref{fig-pg}) to the total QE (Fig.~\ref{fig-pr}) is small ($\sim 5\%$). This follows from the the fact that the photogeneration is a single II event occurring on short ($<100$~fs) dephasing timescale. In contrast, during the population relaxation multiple II events take place on much longer ($\gtrsim 1$~ps) timescale and provide major contribution to QE. 

To conclude, use of IESM parameterized by the effective mass Hamiltonian allowed us to investigate the mechanisms involved in the photogeneration and population relaxation processes contributing to CM in PbSe NCs and bulk. We argue that direct photogeneration has the impact ionization nature and makes weak contribution to the total QE. The use of intensive DOS and effective Coulomb interaction further allowed for a rigorous analysis of the QE scaling with NCs size and comparison with the well defined bulk limit. We have found that QE in NCs plotted on {\it absolute} energy scale does not exceed that in bulk. This is in agreement with reported experimental data and confirms that the quantum-confinement-induced reduction in the biexciton DOS makes dominant contribution to QE. Finally note that the {\it power} performance of photovoltaic devices  depends on the band gap scaling, and even for the case of weak Coulomb enhancement, the NCs are potentially good candidates for such applications. Increase in the effective Coulomb interaction and biexciton DOS through the nanoscale heterostructuring and surface functionalization should further improve their prospective performance.           

This work was supported by the BES Office, DOE, Los Alamos LDRD funds, and CNLS. We thank V. Klimov and D. Smith for stimulating discussions and comments on the manuscript. 


\begin{thebibliography}{10}

\bibitem{kolodinski93}
S. Kolodinski, J. Werner, T. Wittchen, and H. Queisser, Appl.~Phys.~Lett. {\bf
  63},  2405  (1993).

\bibitem{nozik02}
A.~J. Nozik, Physica~E {\bf 14},  115   (2002).

\bibitem{schaller04}
R.~D. Schaller and V.~I. Klimov, Phys.~Rev.~Lett. {\bf 92},  186601  (2004).

\bibitem{ellingson05}
R.~J. Ellingson {\it et~al.}, Nano Lett. {\bf 5},  865   (2005).

\bibitem{schaller05}
R.~D. Schaller, V.~M. Agranovich, and V.~I. Klimov, Nature~Phys. {\bf 1},  189
   (2005).

\bibitem{nair07}
G. Nair and M.~G. Bawendi, Phys.~Rev.~B {\bf 76},  081304(R)  (2007).

\bibitem{nair08}
G. Nair, S.~M. Geyer, L.-Y. Chang, and M.~G. Bawendi, Phys.~Rev.~B {\bf 78},
  125325  (2008).

\bibitem{benlulu08}
M. Ben-Lulu {\it et~al.}, Nano Lett. {\bf 8},  1207  (2008).

\bibitem{pijpers08}
J.~J.~H. Pijpers {\it et~al.}, J.~Phys.~Chem.~C {\bf 112},  4783  (2008).

\bibitem{beard09}
M.~C. Beard {\it et~al.}, Nano Lett. {\bf 9},  836  (2009).

\bibitem{mcguire08}
J.~A. McGuire {\it et~al.}, Acc.~Chem.~Res. {\bf 41},  1810  (2008).

\bibitem{mcguire10}
J.~A. McGuire {\it et~al.}, Nano Lett. {\bf 10},  2049  (2010).

\bibitem{pijpers09}
J.~J.~H. Pijpers {\it et~al.}, Nature~Phys. {\bf 5},  811  (2009).

\bibitem{rupasov07}
V.~I. Rupasov and V.~I. Klimov, Phys.~Rev.~B {\bf 76},  125321  (2007).

\bibitem{shabaev06}
A. Shabaev, A.~L. Efros, and A.~J. Nozik, Nano Lett. {\bf 6},  2856   (2006).

\bibitem{witzel10}
W.~M. Witzel, A. Shabaev, C.~S. Hellberg, V.~L. Jacobs, and A.~L. Efros, 
Phys.~Rev.~Lett. {\bf 105},  137401  (2010).

\bibitem{franceschetti06}
A. Franceschetti, J.~M. An, and A. Zunger, Nano Lett. {\bf 6},  2191   (2006).

\bibitem{rabani08}
E. Rabani and R. Baer, Nano Lett. {\bf 8},  4488  (2008).

\bibitem{allan08}
G. Allan and C. Delerue, Phys.~Rev.~B {\bf 77},  125340  (2008).

\bibitem{prezhdo09}
O.~V. Prezhdo, Acc.~Chem.~Res. {\bf 42},  2005  (2009).

\bibitem{rabani10}
E. Rabani and R. Baer, Chem.~Phys.~Lett. {\bf 496},  227   (2010).

\bibitem{delerue10}
C. Delerue, G. Allan, J.~J.~H. Pijpers, and M. Bonn, Phys.~Rev.~B {\bf 81},
  125306  (2010).

\bibitem{silvestri10}
L. Silvestri and V.~M. Agranovich, Phys.~Rev.~B {\bf 81},  205302  (2010).

\bibitem{piryatinski10}
A. Piryatinski and K.~A. Velizhanin, J.~Chem.~Phys. {\bf 133},  084508  (2010).

\bibitem{beard10}
M.~C. Beard {\it et~al.}, Nano Lett. {\bf 10},  3019  (2010).

\bibitem{kang97}
I. Kang and F.~W. Wise, J.~Opt.~Soc.~Am.~B {\bf 14},  1632   (1997).

\bibitem{bonati07}
C. Bonati {\it et~al.}, Phys.~Rev.~B {\bf 76},  033304  (2007).

\end{thebibliography}

\end{document}